\documentclass[twocolumn, superscriptaddress, floatfix, showpacs, aps, prb]{revtex4}
\usepackage{graphicx}
\usepackage{amsmath}
\usepackage{amssymb}
\usepackage[colorlinks=true, citecolor=blue, urlcolor=blue]{hyperref}
\usepackage{bm}

\DeclareMathOperator{\de}{d\!}

\DeclareMathOperator{\e}{e}
\DeclareMathOperator{\im}{Im}
\newcommand{\abs}[1]{\lvert#1\rvert}

\newcommand{\ket}[1]{|#1\rangle}

\begin{document}

\title{Thermal conductance as a probe of the non-local order parameter for a topological superconductor with gauge fluctuations}

\author{B. van Heck}
\affiliation{Instituut-Lorentz, Universiteit Leiden, P.O. Box 9506, 2300
RA Leiden,  The Netherlands}
\author{E. Cobanera}
\affiliation{Instituut-Lorentz, Universiteit Leiden, P.O. Box 9506, 2300
RA Leiden,  The Netherlands}
\author{J. Ulrich}
\affiliation{Institute for Quantum Information, RWTH Aachen University, 52056 Aachen, 
Germany}
\author{F. Hassler}
\affiliation{Institute for Quantum Information, RWTH Aachen University, 52056 Aachen, 
Germany}

\date{January 2014} 
\begin{abstract} 
We investigate the effect of quantum phase slips on a helical quantum wire
coupled to a superconductor by proximity.  The effective low-energy
description of the wire is that of a Majorana chain minimally coupled to a
dynamical $\mathbb{Z}_2$ gauge field.  Hence the wire emulates a
matter-coupled gauge theory, with fermion parity playing the role of the
gauged global symmetry.  Quantum phase slips lift the ground state degeneracy
associated with unpaired Majorana edge modes at the ends of the chain, a
change that can be understood as a transition between the confined and the
Higgs-mechanism regimes of the gauge theory.  We identify the quantization of
thermal conductance at the transition as a robust experimental feature
separating the two regimes.  We explain this result by establishing a
relation between thermal conductance and the Fredenhagen-Marcu string
order-parameter for confinement in gauge theories.  Our work indicates that
thermal transport could serve as a measure of non-local order parameters for
emergent or simulated topological quantum order.
\end{abstract} 


\maketitle

Topological phases of matter cannot be characterized by any local
order-parameter and, hence, signatures of these phases are not accessible by a
local experimental probe.  For free fermions, the complete classification of
topological phases has recently been established
\cite{Kitaev2009,Ryu2010,Wen2012a} and a connection between the
(experimentally accessible) linear response properties of a system and the
value of its topological invariant has been obtained.  A prominent and
illustrative example are one-dimensional (1D) topological superconductors,
\cite{Kitaev2001,Fu2009,Lutchyn2010,Oreg2010} currently the subject of
intense theoretical \cite{Alicea2012,Beenakker2013} and experimental
investigation.\cite{Mourik2012, Das2012, Deng2012, Rokhinson2012, Knez2012,
Hart2013}  In this case, the topological phase is characterized by unpaired
Majorana zero modes at the ends of the superconductor, whose presence allows
to non-locally store one bit of quantum information encoded in the total
fermion parity of the superconductor.\cite{Kitaev2001} This topological phase
can be recognized by striking transport properties. \cite{Beenakker2013}
Perfect Andreev reflection off a Majorana end mode leads to a quantized
zero-bias conductance of $G_0=2e^2/h$.
\cite{Law2009,Flensberg2010,Wimmer2011,Fidkowski2012} The peak can only be
removed if the system undergoes a phase transition into a phase without
Majorana modes.  Exactly at the transition, the two unpaired Majorana modes
combine into a perfectly transmitting mode.  As a consequence, the thermal
conductance through the wire peaks at a value equal to its superconducting
quantum $K_0=\pi^2 k_B^2 T/6h$ at temperature $T$.\cite{Akhmerov2011} The quantization of the peak is a way to identify the topological phase transition even in a wire of finite size.\cite{Akhmerov2011} In the
topologically trivial phase, both zero-bias Andreev and thermal conductance
are zero.

\begin{figure}[t] 
  \centering
  \includegraphics[width=\linewidth]{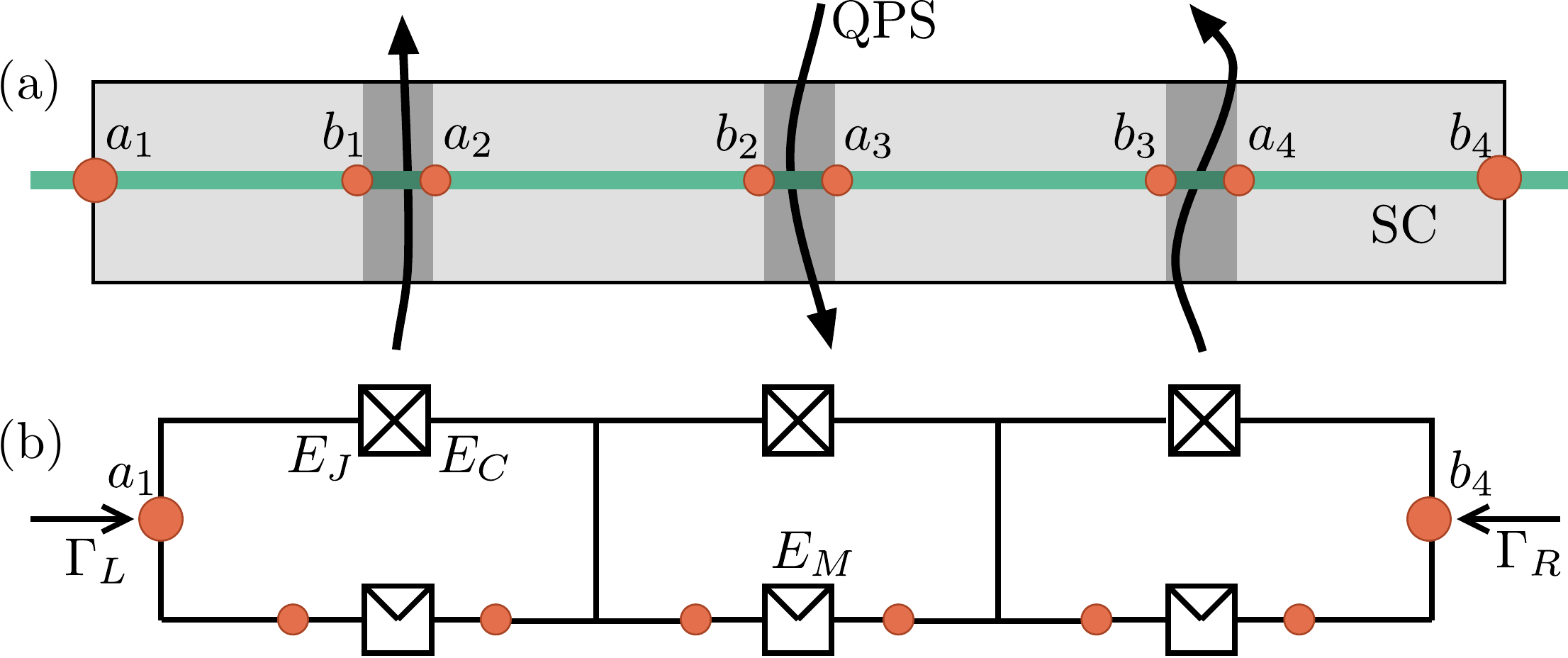}
\caption{%
Panel (a): An $s$-wave superconductor (gray) is deposited on top of a helical
quantum wire (green), which can be for example a semiconducting nanowire or
the edge of a quantum spin Hall insulator.  We consider the effect of quantum
phase slips in the superconductor (black arrows).  Once a moderate magnetic
field is applied to break time reversal invariance, Majorana modes (orange
dots) appear at the ends of the wire and at weak links when the phase slips
happen.  Panel (b): We show an equivalent circuit describing the system [see
Eq.~\eqref{eq:action}].  Here, as usual, a box with a cross denotes a
Josephson junction and its capacitance.  On the other hand, a box with only
half of a cross denotes the $4\pi$-periodic Majorana junction.  Arrows
represent coupling to external leads to the Majorana modes at the end with
strength $\Gamma_L$ and $\Gamma_R$.
}\label{fig:setup} 
\end{figure}

It is currently a challenge in condensed matter physics to extend the
classification of topological phases to interacting fermionic systems (see Refs.~\onlinecite{Fidkowski2010, Wen2013, Hung2013}) and in particular to provide a
similar connection with experimental probes.  Often, insight into interacting
topological phases is offered by non-local order parameters.
\cite{Wegner1971,Nussinov2009} However, such quantities lack an obvious
thermodynamic meaning and do not enable natural mean-field approximations.  If
available, they are useful theoretical tools, \cite{Cobanera2013,Bahri2013}
without direct experimental implications.  Thus, not surprisingly, they are
dubbed `hidden'.

In this paper, we will show that non-local order parameters can be directly linked to transport properties in the linear response regime.  We will show
this for the case of a 1D topological superconductor subject to quantum phase
slips, see Fig.~\ref{fig:setup}.  The system is described by an effective
interacting Hamiltonian akin to a matter-coupled lattice gauge theory, the 1D
$\mathbb{Z}_2$ Higgs model.\cite{Fradkin1979} A non-local order parameter is
in this case known: the Fredenhagen-Marcu string order parameter,
\cite{Fredenhagen1983} originally proposed as a criterion for confinement
\cite{Fredenhagen1986} and recently revisited in the context of topological
order.\cite{Gregor2011} We will show that the Fredenhagen-Marcu order
parameter is connected in our system to a simple transport coefficient, the
thermal conductance.

Let us start by discussing the role of quantum phase slips (QPS) in topological
superconductors. QPS are quantum tunneling events where
the phase of the superconducting order parameter changes locally by $2\pi$.
In 1D, QPS destroy the superconducting phase at zero temperature
\cite{Zaikin1997,Lau2001,Golubev2001} and thus remove the topological
protection of a Majorana qubit,\cite{Sau2011} since the latter presupposes
the superconducting ordering which breaks the electromagnetic U(1) symmetry
down to $\mathbb{Z}_2$.  For $d$-dimensional superconductors with $d>1$, QPS
are suppressed as they generate a domain wall in the superconducting order
parameter, leading to $\kappa \propto \exp[-(L/\xi)^{d-1}]$, with $L$ the
linear dimension of the system and $\xi$ the coherence length.  In this sense,
Kitaev's model of topological protection is not purely one-dimensional, since
a bulk (3D) superconductor is crucial for achieving the fault-tolerance of a
Majorana qubit.\cite{Sau2011}

To study QPS in a concrete setting, we follow the approach of
Ref.~\onlinecite{Matveev2002} and consider a chain of coupled superconducting
islands, with superconducting phase $\phi_m$, placed on top of a nanowire or
of a quantum spin Hall edge, see Fig.~\ref{fig:setup}.  The junctions between
the islands then naturally form weak links through which QPS happen.  The
Euclidean action describing a chain of $N$ islands reads
$\mathcal{S}=\int_0^{1/T} \mathcal{L}\de t $, with
\cite{Pekker2013,Pekker2013b}
\begin{align}\notag
  \mathcal{L}&=\sum_{m=1}^{N-1}
\biggl[\frac{\dot\varphi^2_m}{2E_C}+E_J(1-\cos \varphi_m)\\
&\qquad-i E_M b_ma_{m+1}\cos(\varphi_m/2) \biggr]- \sum_{m=1}^{N}
iha_mb_m\,;\label{eq:action}
\end{align}
here, $\varphi_m= \phi_{m+1} - \phi_m$ is the phase difference across each
junction.  The charging energy $E_C = e^2/2C$ and the Josephson energy $E_J =
\hbar I_c/2e$ are respectively determined by the capacitance $C$ and the
critical current $I_c$ of the junction.  A topological superconducting wire
hosts two Majorana zero-energy modes $a_m,b_m$ on each island.  They are
responsible for the term proportional to $E_M$ in \eqref{eq:action},
describing tunneling of individual electrons \cite{Fu2009}.  The Hermitian
operators $a_m, b_m$ obey the anti-commutation rules $\{a_m, b_n\}=0$ and
$\{a_m, a_n\}=\{b_m,b_n\}=2\delta_{mn}$.  Additionally, the finite size of the
islands leads to an overlap between Majorana modes and an associated energy
splitting denoted by $h$.  The local fermion parity $p_m=\pm 1$ at each
junction is defined via the occupation number of a fermionic mode
$c_m=\tfrac{1}{2}\left(b_m-i a_{m+1}\right)$ as
$p_m=1-2c^\dagger_mc_m=ib_ma_{m+1}$.  The total fermion parity operator
$(-1)^F=ia_1\,\prod_{m=1}^{N-1} p_m\,b_N$ is a global symmetry of the system.

Different from previous studies \cite{Shivamoggi2010,Xu2010,Hassler2012}, we
are interested in the regime $E_J\gg E_M, E_C, h$, where the superconducting
phase difference at any junction can only be a multiple of $2\pi$, due to the
large Josephson energy.  The relevant quantum fluctuations in the chain are
QPS connecting classical minima, whose amplitude $\kappa \simeq
(E_CE_J^3)^{1/4}\,\exp(-8\sqrt{E_J/E_C})$ can be computed in the semiclassical
approximation.\cite{Matveev2002} A shift of $\varphi_m$ by $2\pi$ changes
the sign of $E_M\cos(\varphi_m/2)$ and thus it also changes the
energetically-favored value of the junction parity $p_m$.\cite{Pekker2013} In this regime, the value of $\cos(\varphi_m/2)$ is reduced to a
$\mathbb{Z}_2$ quantum degree of freedom.  The effective Hamiltonian of the
chain,
\begin{equation}\label{eq:Ham}
H=-\sum_{m=1}^N iha_mb_m-\sum_{m=1}^{N-1} \Bigl[ iE_M b_ma_{m+1}\tau^z_m+
\kappa \tau^x_m \Bigr] \, ,
\end{equation}
describes Majorana modes coupled to $N-1$ Pauli matrices $\tau_m^z
=\cos(\varphi_m/2)$, one per junction.\cite{Cobanera2013} The last term in
the Hamiltonian describes QPS that change $\varphi_m$ by $2\pi$ at a rate
$\kappa/\hbar$.

In the absence of fluctuations of the superconducting order parameter, that
is, at $\kappa=0$, we recover the Kitaev model.  In this case, the $\tau^z_m$
degrees of freedom are redundant.  The Hamiltonian $H$ can be
block-diagonalized by freezing them in some classical configuration.  All
blocks in this decomposition have identical energy spectra.  For any classical
configuration of the spins $\tau^z_m$, a quantum critical point at $h=E_M$
separates a topologically non-trivial phase at $h<E_M$ from the trivial phase
at $h>E_M$.  The non-trivial phase has a twofold ground state degeneracy if
both even and odd total fermion parities $(-1)^F = \pm 1$ are considered,
signaling the presence of unpaired Majorana modes at either ends of the
chain.

The interaction of the fluctuating superconducting phase with the Majorana
modes is such that, for each island, a local symmetry $C_m$ of $H$ emerges,
given by
\begin{align}
C_1 &= ia_1b_1\,\tau^x_1,\qquad C_{N}=\tau^x_{N-1}\, ia_{N}b_{N}, \nonumber\\
C_{m}&=\tau^x_{m-1}\,ia_{m}b_{m}\,\tau^x_{m} \quad(m=2,\dots,N-1)\,.\label{eq:symmetries}
\end{align}
These local symmetries are gauge symmetries and appear because the phase
difference and fermion parity of a junction are not independent degrees of
freedom \cite{Fu2010}: a change in the occupation number of the fermionic mode
$c_m$ is equivalent to advancing the phase $\varphi_m$ by $2\pi$.  As a
result, the global fermion parity $(-1)^F$ can be expressed as a product of
the local gauge-symmetries $(-1)^F=\prod_{m=1}^NC_m$.\cite{Cobanera2013} It
follows that the $\tau^z_m$ play the role of a $\mathbb{Z}_2$ gauge field,
minimally coupled to the fermionic degrees of freedom and with dynamics
generated by QPS.

The link to lattice field theories can be made more explicit.  Our effective
Hamiltonian $H$ of Eq.~\eqref{eq:Ham} can be interpreted as an approximation
to the lattice-regularized 1D Higgs model,\cite{Fradkin1979} given by
\[
  H_\text{H}= 
 - \sum_{m=1}^{N} \frac{\lambda}{2} \partial^2_{\phi_m} 
 +  \sum_{m=1}^{N-1} \Bigl[ -\frac{g^2}{2}\partial_{\theta_m}^2
  +  v^2\cos(\varphi_{m}-\theta_m) \Bigr]\ .
\]
This Hamiltonian follows by standard techniques \cite{Wilson1974,Kogut1975}
from the Euclidean action of the Higgs model of Ref.~\onlinecite{Fradkin1979}.
Here, the angular variables $\phi_m, \theta_m$ represent the Higgs and
electromagnetic gauge field respectively.  The parameter $v^2$ is the vacuum
expectation value of the Higgs field in the broken-symmetry state.  The
parameters $\lambda, g^2$ control the strength of the fluctuations of the
matter and gauge fields.

Our Hamiltonian $H$ is obtained from that of the Higgs model $H_\text{H}$ by
using the approximation $-\pi^2\partial_x^2/2 \approx \cos(\pi \partial_x) -1$
for $x= \phi_m, \theta_m$ and truncating the angular variables to the values
$\phi_m, \theta_m \in \{0,\pi\}$.  Within the truncated Hilbert space,
$\cos(\pi \partial_{\phi_m}) =\sigma^x_m$ and $\cos(\pi \partial_{\theta_m}) =
\tau^x_m$.  Hence, $H_\text{H}$ reduces (up to an irrelevant additive
constant) to the spin chain Hamiltonian
\begin{equation}\label{eq:Hamdual}
  H_{\mathbb{Z}_2} = \sum_{m=1}^N \frac{\lambda}{\pi^2}\sigma^x_m
  +\sum_{m=1}^{N-1} \Bigl[ \frac{g^2}{\pi^2}\tau^x_m+
  v^2 \sigma^z_m\tau^z_m\sigma^z_{m+1} \Bigr].
\end{equation}
The Hamiltonian $H_{\mathbb{Z}_2}$ is precisely that of the $\mathbb{Z}_2$
Higgs model.\cite{Fradkin1979} Finally, the Jordan-Wigner transformation
$a_m=\sigma^x_m\prod_{j=1}^{m-1}\sigma^z_j $,
$b_m=\sigma^y_m\prod_{j=1}^{m-1}\sigma^z_j$ shows that the $\mathbb{Z}_2$
Higgs model is equivalent to our Hamiltonian $H$, provided we identify $h=
-\lambda/\pi^2$, $\kappa= - g^2/\pi^2$, and $E_M=-v^2$.\cite{Note1}

As our effective Hamiltonian $H$ is related to the Higgs model, we might
expect the Higgs mechanism to be present.  As a result, gapless excitations
should become gapped for arbitrarily small values of $\kappa$, that is, for
arbitrarily weak fluctuations of the superconducting order parameter.  In
other words, the small but finite charging energy $E_C$ of each island breaks
the ground state degeneracy and splits the otherwise unpaired Majorana modes.
In this way, the Higgs-mechanism offers a way to locally break the topological
degeneracy of the Majorana chain.  It is known that this expectation is indeed
correct in the thermodynamic limit, as at $\kappa\neq 0$ the Hamiltonian
\eqref{eq:Hamdual} has no phase transitions and describes a gapped phase with
a single ground state.\cite{Fradkin1979,Uzelac1980,Cobanera2013} However, in
a finite chain signatures of the topologically non-trivial phase, which is
present at $\kappa=0$ and $h<E_M$, should survive up to a finite value of
$\kappa$. If this is true, then the Hamiltonian of a finite chain should be gapless along a line in the $(h,\kappa)$ plane.

In the following, we will show that in the linear response regime, the topological
transition reflects itself in the thermal conductance $K$ through the system
also at finite $\kappa$, whereas upon increasing $\kappa$, the local probe of
Andreev conductance $G$ quickly becomes blind to it.  To this end, we couple
the left and right end of the chain with Hamiltonian $H$ to normal leads
through tunneling Hamiltonians \cite{Fu2010}
$H_\textrm{L}=\gamma_\textrm{L}c^\dagger_\textrm{L}
a_1\e^{-i\phi_1/2}+\text{H.c.}$,
$H_\textrm{R}=\gamma_\textrm{R}c^\dagger_\textrm{R}
b_N\e^{-i\phi_N/2}+\text{H.c.}$; here, $\gamma_\textrm{L}$ and
$\gamma_\text{R}$ denote the amplitudes for tunneling events into the left (L)
and right (R) leads, and $c^\dagger_{\textrm{L}}, c^\dagger_{\textrm{R}}$ are
the creation operators for electrons in the non-interacting leads.  We fix the
gauge by choosing $\phi_1=0$, so that $\phi_N=\sum_{m=1}^{N-1}\varphi_m$.  In
the low-energy limit, $\e^{-i\phi_N/2}=\prod_{m=1}^{N-1}\tau^z_m$, so we get
\begin{equation}
H_\textrm{L}=\gamma_\textrm{L}(c_\textrm{L}^\dagger+c_\textrm{L})a_1\,,\;
H_\textrm{R}= 
\gamma_\textrm{R}(c^\dagger_\textrm{R}+c_\textrm{R})b_N\prod_{m=1}^{N-1}\tau^z_m
\,.
\end{equation}
The tunneling Hamiltonians must break one of the gauge symmetries, since a
tunneling event changes the total fermion parity. Due to our
gauge choice, we obtain  $\{H_\textrm{L,R},C_1\}=0$ while 
$[H_\textrm{L,R},C_m]=0$ for $m=2,\dots,N$.

\begin{figure*}[tb]
\centering\includegraphics[width=0.9\linewidth]{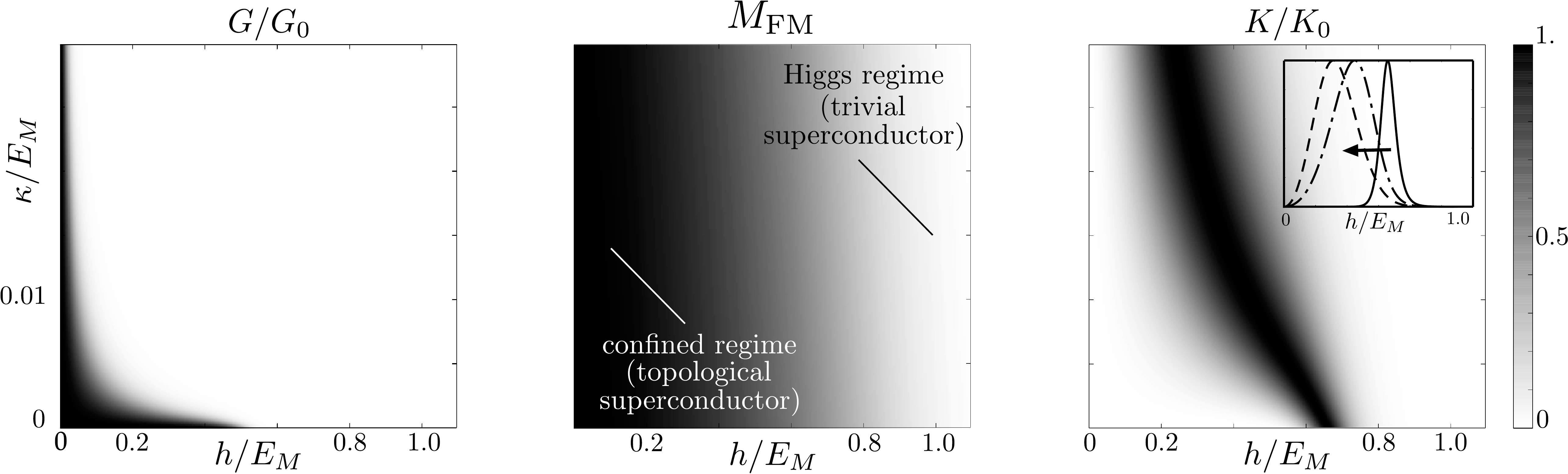}
\caption{Numerical results: Andreev conductance (left, in units of
  $G_0=2e^2/h$), Fredenhagen-Marcu order parameter $M_\textrm{MF}$ (center),
  and thermal conductance (right, in units of $K_0=\pi^2 k_B^2 T/6h$) plotted
  as a function of $h$ and $\kappa$.  The results are obtained for a chain of
  $N=11$ islands with coupling constant $\Gamma=0.01E_M$ to the leads, in the
  limit of vanishing temperature $T$ and small applied voltage $V$.  The
  Andreev conductance $G$ is averaged over a small voltage interval
  $10^{-4}E_M$ to account for the finite-size energy splitting of the Majorana
  modes as for a finite size wire we trivially have $G=0$.\cite{Flensberg2010,Hassler2012} The Andreev conductance only shows a
  signal along the axis.  On the other hand, the string-order parameter allows
  to distinguish a confined regime - corresponding to the topological regime with Majorana end modes - from
  the trivial Higgs-regime.  The separation between these two regimes can be clearly
  identified by the peak in the thermal conductance. The inset in the right panel shows line cuts of the thermal conductance at $\kappa/E_M=0, 0.01, 0.02$, going from right to left as shown by the arrow. Due to finite size
  effects, the transition at $\kappa=0$ is shifted from $h/E_M=1$ to $h/E_M\simeq
  (\Gamma/E_M)^{1/11}\simeq 0.7$, as expected.  Increasing $\kappa$, the
  topological regime shrinks and only the Higgs-regime survives.}
  \label{fig:numerics}
\end{figure*}

The Andreev conductance $G$ is determined by the charge transport across a
normal metal-superconductor interface.  To compute $G$, we set
$\gamma_\text{R}=0$, $\gamma_\text{L}= \gamma$, apply a bias voltage $V$ to
the left lead, and ground the rightmost superconducting island.  In contrast,
the thermal conductance $K$ is determined by the heat transport between two
normal leads.  To compute $K$ we set $\gamma_\text{L}= \gamma_\text{R} =
\gamma$ and establish a small temperature difference between the right lead at
temperature $T$ and the left lead at temperature $T+\delta T$.  In the limit
$T,V\to 0$, we obtain $G$ as $G=G_0\,\Gamma\im[G_{11}(0)]$,\cite{Hutzen2012}
 and $K$ as $K=4K_0\Gamma^2\,\abs{G_{1N}(0)}^2$,\cite{Note2} in terms of the
tunnel coupling $\Gamma = 2 \pi \abs{\gamma}^2 \rho_0$ to a wide-band lead with
density of states $\rho_0$ and the retarded Green's functions
\begin{align}
 G_{11}(\omega) &= -i \int_0^\infty \!\de t \,e^{i \omega t}
\Bigl\langle\{a_1(t), a_1(0)\}\Bigr\rangle\,, \label{g1}\\
G_{1N}(\omega) & = -i \int_0^\infty \!\de t\, e^{i\omega t}\Bigl\langle\{b_N(t)\,\textstyle\prod_{m=1}^{N-1}
\tau^z_m(t),a_1(0)\}\Bigr\rangle\, . \nonumber
\end{align}
The averages in Eqs.~\eqref{g1} are taken over the ground state wave function
$|0\rangle$ of our effective Hamiltonian $H$.  For any $\kappa,h,E_M>0$, the
ground state of $H$ is unique and belongs to the gauge-invariant sector with
$C_m |0\rangle= |0\rangle$ for all $m$.\cite{Note3} The time-evolution in
Eqs.~\eqref{g1} is determined by the total Hamiltonian
$H_\textrm{tot}=H+H_\textrm{R}+H_\textrm{L}$.  The retarded Green's function
$G_{11}(t) = \int\!(\de \omega/2\pi)\,e^{-i \omega t} G_{11}(\omega)$ is the
amplitude for a reflection process whereby an electron enters the chain from
the left lead at time $t_i=0$ and exit again from the left lead after a time
$t_f=t$.  Similarly, $G_{1N}(t)$ is the amplitude for a transmission process
whereby the electron enters at $t_i=0$ from the left lead and exits from the
right lead after a time $t_f=t$.

We highlight that the thermal transport probes \emph{non-local} quasiparticle
transfer processes through the chain characterized by the string correlator
$G_{1N}(t)=-i\langle\{b_N(t)\,\textstyle\prod_{m=1}^{N-1}\tau^z_m(t),a_1(0)\}
\rangle$, which is a generalization of the conventional correlator
$-i\langle\{b_N(t),a_1(0)\}\rangle $ studied in the context of the Majorana
chain without the gauge degrees of freedom.\cite{Flensberg2010} Due to the presence of the gauge string $\prod_{m=1}^{N-1}\tau^z_m$, the Green's function $G_{1N}$ is similar to the Fredenhagen-Marcu string-order parameter,\cite{Fredenhagen1983}
\begin{equation}
M_\text{FM}=-i\langle b_N\, \textstyle\prod_{m=1}^{N-1}\tau^z_m\, a_1\rangle,
\end{equation}
which measures the presence of the topological phase in the model with
fluctuation gauge degrees of freedom. In the following we probe this relation numerically.

To calculate the Green's function $G_{mn}(\omega)$, we follow the approach
\cite{Lacroix1981} of decoupling the gauged Majorana chain from the leads to
first order in the lead coupling $\Gamma$ and neglecting higher-order
(co-)tunneling processes.  The bare Green's functions without the leads are
calculated by exact diagonalization of the Hamiltonian \eqref{eq:Ham}, using a
Lehmann spectral representation in terms of the exact eigenstates.  The
presence of the symmetries \eqref{eq:symmetries} greatly simplifies the task
of computing $G_{mn}$.  In fact, we only need to know the energy and the
wavefunction of the ground state $\ket{0}$ and of all the states $\ket{\psi}$
such that $C_1\ket{\psi}=-\ket{\psi}$ while $C_{m\neq1}\ket{\psi}=\ket{\psi}$.
Indeed, since $C_1$ is the only symmetry of $H$ which does not commute with
the tunneling Hamiltonian $H_\textrm{L,R}$, but anti-commutes instead, these
are the only excited states to which transitions from the ground states are
possible upon tunneling of an electron from the leads.  For a chain of $N$
islands, there are $2^{N-1}$ of these states---against a total Hilbert space
dimension of $2^{2N-1}$.\cite{Note4}

The numerical results for a chain of $N=11$ islands are shown in
Fig.~\ref{fig:numerics}.  At $\kappa=0$, coherently with known results, we
observe an Andreev reflection plateau at $G_0$ in the non-trivial regime and a
thermal conductance peak at the transition, which appears shifted to $h
\simeq E_M(\Gamma/E_M)^{1/N}$ due to finite chain size and coupling to the
leads.  At finite $\kappa$, the Andreev plateau is quickly suppressed, except
close to $h=0$, a limit where two isolated Majorana modes are always present.
However, the quantized peak in thermal conductance persists in the interacting
part of the parameter space, indicating the presence of a gapless transmitting
mode and hence a strong signature of the existence of a topological regime.  In
fact, the position of the thermal conductance peak qualitatively follows the
line of maximum change in the order parameter. We have checked that the agreement persists when varying the system size $N$.

To conclude, we have shown that QPS in a Majorana chain implement the
$\mathbb{Z}_2$ Higgs model where the fluctuations of the gauge field are
determined by the rate $\kappa/\hbar$ for QPS.  QPS \emph{locally} destroy the
topological phase of the Kitaev model at fixed fermion parity via a
$\mathbb{Z}_2$ version of the Higgs mechanism.  However, for finite system
size and small $\kappa$, signatures of the topological phase remain visible in
the thermal conductance through the system.  The reason is that it is linked
to the Fredenhagen-Marcu order parameter for the $\mathbb{Z}_2$ Higgs theory,
which indicates the topological regime with gauge fluctuations present.
The thermal conductance provides a clear transport signature of the transition
from the topological to the trivial regimes in the presence of the interactions with the gauge field,
whereas no signature of the transition is present in the Andreev conductance
at a finite rate of QPS.  Our results suggest that in topological quantum
matter, bulk transport measurements offer access to non-local order
parameters, just like susceptibility measurements do for local order
parameters in broken-symmetry phases.  It remains an interesting question for
further studies how this scenario can be generalized to higher dimensions and
non-Abelian gauge fields.

\emph{Acknowledgments.} We thank Anton Akhmerov, Carlo Beenakker, Gerardo Ortiz 
and Dirk Schuricht for fruitful discussions. BvH and EC were supported by the 
Dutch Foundation for Fundamental Research on Matter (FOM), the Netherlands 
Organization for Scientific Research (NWO/OCW), and an ERC Synergy Grant. 
JU and FH are grateful for support from the Alexander von Humboldt foundation and from the RWTH Aachen Seed Funds.

\end{document}